# Chatbot-supported Thesis Writing:
# An Autoethnographic Report

Nicolas Schwenke[1][0009-0007-7371-3860], Heinrich Söbke[1,2][0000-0002-0105-3126], and Eckhard Kraft[1][0000-0002-7775-4551]

[1] Bauhaus-Institute for Infrastructure Solutions (b.is), Bauhaus-Universität Weimar,
Goetheplatz 7/8, 99423 Weimar, Germany
{nicolas.schwenke|heinrich.soebke|eckhard.kraft}@uni-weimar.de
[2] Hochschule Weserbergland, Am Stockhof 2, 38175 Hameln, Germany
soebke@hsw-hameln.de

**Abstract.** The release of the large language model based chatbot ChatGPT in November 2022 has brought considerable attention to the subject of artificial intelligence, not only in the public. From the perspective of higher education, ChatGPT challenges various learning and assessment formats as it significantly reduces the effectiveness of their learning and assessment functionalities. In particular, ChatGPT might be applied to formats that require learners to generate text, such as bachelor theses or student research papers. Accordingly, the research question arises to what extent writing of bachelor theses is still a valid learning and assessment format. Correspondingly, in this study, the first author was asked to write his bachelor's thesis exploiting ChatGPT. For tracing the impact of ChatGPT, methodically an autoethnographic approach was used. First, all considerations on the potential use of ChatGPT were documented in logs and secondly, all ChatGPT chats were logged. Both logs and chat histories were analyzed and are presented along to the recommendations for students regarding the use of ChatGPT suggested by Gimpel et al. (2023). In conclusion, ChatGPT is beneficial in thesis writing during various activities, such as brainstorming, structuring and text revision. However, there arise limitations, e.g., in referencing. Thus, ChatGPT requires a continuous validation of the outcomes generated fostering learning. Currently, ChatGPT is to be valued as a beneficial tool in thesis writing. However, writing a conclusive thesis still requires the learner's meaningful engagement. Accordingly, writing a thesis is still a valid learning and assessment format. With further releases of ChatGPT, an increase in capabilities is to be expected and the research question needs to be reevaluated from time to time.

## 1    Introduction

In November 2022, the large language model (LLM)-based chatbot ChatGPT was released and received a lot of public attention due to its enormous capabilities. ChatGPT is an example of increasingly powerful artificial intelligence (AI) tools that allow opening up new fields of application (Schmoll et al., 2020). Higher education (HE), with numerous teaching and assessment formats, also provides numerous opportunities for AI tools. For example, ChatGPT may be implemented in new teaching formats that specifically align the learning objectives to the student's capabilities and support individual learning processes with accompanying personalized feedback (Alexander et al., 2019; Zawacki-Richter et al., 2019). Additionally, the use of AI tools may have a disruptive impact on assessment formats in HE, as the learning and assessment functionality of the formats is at least reduced, if not lost (de Witt & Rampelt, 2020; Gao et al., 2022; King & ChatGPT, 2023; Susnjak, 2022; Zhai, 2022). For instance, if an essay is written entirely by ChatGPT, the student's writing skills will not be encouraged to the level intended. Also, when answers to exam questions are generated by students using ChatGPT, there is an assessment of ChatGPT's "knowledge" rather than of the student's understanding. In conclusion, there is a demand for adaptation of existing teaching and assessment formats to the capability of current AI tools. In addition, organizational guidelines must ensure responsible use of AI in HE (Gimpel et al., 2023).

As a chatbot, ChatGPT is well-suited in those teaching and assessment formats mostly based on generating texts. Such a format also includes theses, such as bachelor theses and student research papers. For theses, the use of ChatGPT cannot be excluded in the same way as for, e.g., proctored, or oral exams. Accordingly, it is important to answer the research question regarding the impact ChatGPT might have on assessment and learning functionality of theses. Thus, the first author was assigned to do his bachelor thesis using ChatGPT wherever possible and meaningful. Based on an autoethnographical approach, the impact of the AI tool on text-heavy formats of HE is to be identified. In accordance, the next section presents the state of the literature leading from AI in education in general towards the writing of theses with the help of LLM-based AI tools. Section 3 describes the methods and also the general findings of the bachelor thesis. The autoethnographical results are described in section 4. Finally, in the subsequent two sections, the results are discussed, and the conclusions are drawn.

## 2      Literature Review

Particularly in education, new "artificial intelligence tools for education (AIEd)" (Baker & Smith, 2019, p. 11) are being pushed forward in  a dedicated field of research (Luckin et al., 2016). The primary tasks of AIEd are to provide support in HE (Baker & Smith, 2019; Luckin et al., 2016), as well as to teach "digital literacy" (Wannemacher & Bodmann, 2021, p. 32). In the following we discuss cornerstones in the field.

### 2.1     AIEd Classification

AIEd may be classified by different approaches. For example, de Witt et al. (2020) describe the relevance of AIEd based on the importance of learning analytics (LA) in the purposeful optimization of everyday HE life. This involves using data collection and analysis to advance to a higher level of comprehension of teaching and learning. Subsequently, based on the interaction between user and AI tools, the respective capabilities should be combined for exploiting the potentials of AIEd.

Based on their systematic literature review, Zawacki-Richter et al. (2019) classify AIEd according to the most frequently named fields of application in everyday HE life. They distinguish between applications that relate to the administrative level of HE, direct use within teaching formats, the organization of studies, and evaluation and monitoring. Contrary to this, Backer & Smith (2019) refer to a classification of AIEd based on system-, teaching- and learning-oriented applications. Thereby, system-oriented applications focus on the provision of efficient infrastructure in study organization and administration (Baker & Smith, 2019). This includes automation of enrollment and admission processes (Zawacki-Richter et al., 2019), as well as automated identification and analysis of data regarding teaching and learning (Wannemacher & Bodmann, 2021). Teaching-oriented applications focus on instructors; accordingly, new teaching formats may be developed that contribute to an individualization of teaching (Baker & Smith, 2019; Luckin et al., 2016). Thus, lectures might evolve into a "consultation meeting" in which instructors interact with learners by providing "advice and assistance" (Handke, 2018, p. 262). This is flanked by the changing of the instructor's role to that of a learning facilitator through the partial takeover of the teaching by AIEd (Digitalisierung, 2016; Handke, 2018; Luckin et al., 2016). Learning-oriented applications, on the other hand, focus on the individual needs of learners. Thereupon, the focus is on adapting learning objectives to the competencies of the instructors by providing location- and time-independent support with personalized feedback (Alexander et al., 2019; Hobert & Berens, 2019; Luckin et al., 2016; Zawacki-Richter et al., 2019). In addition, AIEd can be used specifically to assist with exam preparation, saving significant time for the learner and maximizing their academic success (Gao et al., 2022; Susnjak, 2022; Zhai, 2022).

## 2.2    Acceptance

Besides accessibility and identification of suitable starting points for AIEd, acceptance by learners is a key factor. Stützer (2022) investigated the acceptance of AIEd at four German universities. With a survey, interdisciplinary influences, already tapped individual experiences in interacting with AIEd and further support provided by AI were considered. The results showed that most learners considered the interdisciplinary use of AIEd as a possibility for the integration of digital technologies in HE. In this regard, a correlation was found between acceptance and the research field of the survey participants. Members of scientific study programs showed a higher level of acceptance. In particular, dialog and assistance systems were appreciated by the students. With regard to the risks of using AIEd, opinions were diverse, ranging from no concerns to uncertainty due to unpredictable side effects. Also, in line with the literature (Handke, 2018; Luckin et al., 2016; Qadir, 2022), students preferred the accompanying contact with instructors. Stützer (2022) states that the acceptance of AI tools increases with purposeful use in addition to performance and usefulness.

## 2.3    Influencing Factors

Two influencing factors for AIEd in particular can be identified: First, the COVID-19 pandemic and ongoing digitalization efforts at universities have encouraged the digitization of teaching (Stützer, 2022). Also, learners are accessing external technologies on their own when digital infrastructure is not provided. In some cases, provided HE institutions resources were rejected (Handke, 2018; Susnjak, 2022). In this context, AIEd are also increasingly used outside of any control and an environment that is not didactically visible (Lensing, 2020). On the other hand, the progressive development of AI adapted to teaching (Schmoll et al., 2020) creates access to a wide range of assistance systems, which are increasingly actively integrated into HE (Shirouzu, 2018).

## 2.4    Potentials of LLMs in HE

ChatGPT is based on an LLM. LLMs are a subfield of generative AI and are capable of understanding, processing, and generating natural language. Based on Baker & Smith (2019), starting points of LLM in HE are described in the following. Besides identifying application areas, capabilities of ChatGPT are also discussed. Overall, LLMs may be used qualitatively in HE (Handke, 2018) and contribute to the elevation of the educational potential of learners (Fürst, 2020, p. 342) by individualizing and personalizing learning (Alexander et al., 2019; Baker & Smith, 2019; Zawacki-Richter et al., 2019).

**System-oriented Potentials**. The system-oriented use of LLMs is based on improving organizational conditions through automated data collection (Kieslich et al., 2019), for example, by taking over routine student administration tasks (Alshater, 2022; Baker & Smith, 2019). In addition, LLMs may act as facilitators of communication between students and administrators (Alshater, 2022; Zawacki-Richter et al., 2019). On the other hand, as a universal tool, LLMs may take over the functions of, e.g., search engines and language translators and – in general – facilitate everyday work (Aljanabi et al., 2023).

**Teaching-oriented Potentials.** Students may be supported within lectures by intensifying interactions during learning (Rudolph et al., 2023). In addition, more discussions may be encouraged, and more feedback may be integrated as a supportive element in teaching (Rudolph et al., 2023). Beyond the course, LLMs may contribute to new teaching methods (Qadir, 2022; Rudolph et al., 2023). Further, LLM may support instructors' research activities away from teaching, e.g., as a tool for literature reviews (Aydın & Karaarslan, 2022) and support the preparation of scientific papers (Gao et al., 2022; Qadir, 2022; Susnjak, 2022; Zhai, 2022).

**Learning-oriented Potentials.** In HE, LLMs focus on "Intelligent Tutoring Systems" (ITS) (Zawacki-Richter et al., 2019), which develop new topics interactively with learners and react flexibly to their needs (Azaria, 2022; Susnjak, 2022). In this regard, LLM are to be understood as time- and location-independent tools that enrich learners' daily HE lives flexibly (Frieder et al., 2023; B. Guo et al., 2023). Continuing, LLMs emphasize an experimental approach to experiences in contrast to conventional methods (Rudolph et al., 2023). By intensifying the interaction between learners and digital tools, learning outcomes are improved (Qadir, 2022). On the other hand, LLMs may be used in assessments (OpenAI, 2023) and simulations of exams (Zhai, 2022). Here, the focus is on the preparation of text-oriented coursework, such as essays and student research papers (Aljanabi et al., 2023; OpenAI ChatGPT Generated Results & Ventayen, 2023; Qadir, 2022; Rudolph et al., 2023; Susnjak, 2022; Yeadon et al., 2022; Zhai, 2022).

## 2.5    Integration into Everyday HE Life

The availability of AI tools, such as LLMs (OpenAI, 2023), in HE teaching (Schmoll et al., 2020). , creates the necessity to integrate them in HE. Thus, numerous studies already presented the impact of AI tools on teaching and learning using pilot implementations (Keller et al., 2019; Lensing, 2020; Pedro et al., 2019; Seufert et al., 2020; Zawacki-Richter et al., 2019). An integration of LLMs in HE teaching might comprise the following steps:

**Perception.** A prior requirement for the use of LLMs in HE is an awareness of LLMs' potentials as well as the willingness to changes (Stützer, 2022)**.** Important are opportunities that allow for independent application of LLMs in individual learning environments and subsequent reflection (Stützer, 2022)**.** According to Witt & Rampelt (2020, p. 10), the avoidance of "technology determinism in teaching and learning" is the foundation for exploiting potentials in line with HE goals. Accordingly, an identification of available resources (Digitalisierung, 2016; Kreutzer & Sirrenberg, 2019) and possible applications of LLMs is required for implementation (de Witt & Rampelt, 2020; Gimpel et al., 2023).Guidelines for handling LLM-generated content are required, e.g., how LLM-generated content is declared and what monitoring methods are available on the part of teaching (Gimpel et al., 2023).

**Accumulation.** The second step is the accumulation of LLM usage in teaching. For the most part, digital technologies, including LLMs, are currently not actively integrated into teaching, but only insufficiently interspersed (Digitalisierung, 2016). Pilot deployments of LLM are needed to evaluate to measure performance in actual teaching as well (Wannemacher et al., 2016). Specifically, no coordination and adjustments should be made (Wannemacher et al., 2016)**,** for preventing a "devaluation of classroom teaching" (Handke, 2018, p. 249), which insufficiently takes advantage of the potentials.

**Integration.** The third step integration builds on the alignment of teaching formats with LLMs (Wannemacher & Bodmann, 2021). This may lead to new formats that are not only supplemented with LLMs, but actively use their potential for teaching (Wannemacher & Bodmann, 2021). As a result, teaching formats also serve to deepen what has been learned (Wannemacher et al., 2016) in which instructors moderate the teaching content and assist learners within their individual learning phase in applying LLMs (Alexander et al., 2019). A key aspect is reaching acceptance towards LLMs (Kieslich et al., 2019; Stützer, 2022). Accordingly, a purposeful integration of LLMs that unites the views of instructors and learners might be beneficial (Blumentritt et al., 2020).

## 2.6 Challenges

The meaningful use of LLM in HE is being faced not only with technical difficulties but also with didactic and organizational challenges (Rensing, 2020). Consequently, problems such as discrimination by LLMs in the sense of manipulation of data sets (Dunkelau & Leuschel, 2020) highlight the need for legal frameworks and data protection regulations in handling LLMs (Digitalisierung, 2016; Massmann & Hofstetter, 2020). Further, ensuring inclusive use of LLMs in education is gaining importance (Pedro et al., 2019). The inclusive use of LLM requires the alignment of monetary, ethical, legal, social, and didactic factors (Buxmann & Schmidt, 2021). In addition, LLM-specific literacy skills must be ensured (Büching et al., 2019), and the responsibility of users (Spannagel, 2023) must be emphasized. Building on this, new assessment regulations that consider the use of LLMs in teaching formats (Baker & Smith, 2019; Gimpel et al., 2023) should be integrated.

## 2.7 Curricular Implications

For the integration of LLM into HE, embedding LLM-specific literacy meta skills as learning objectives into curricula is mandatory (de Witt & Rampelt, 2020; Wannemacher & Bodmann, 2021). Accordingly, a curriculum analysis may identify modules suitable for the integration of LLMs (Wannemacher & Bodmann, 2021). Building on the integration in existing modules, new modules may then be provided to teach LLM-specific meta skills (Digitalisierung, 2016).

## 2.8 Chatbots

With the advent of ChatGPT at the latest, chatbots have become a component to be considered in higher education. For example, Wu et al. (2023) conclude from a meta-analysis that chatbots have a large significant effect on learning outcomes, especially in HE and short-running learning activities. Accordingly, Guo et al. (2023) use the example of chatbot-assisted in-class debates to show that chatbots generate new learning activities. Hwang & Chang (2023), however, find that chatbots have been predominantly used in guided learning activities and that more diverse instructional design integration is needed. In a systematic scoping review, Yan et al. (2023) identify nine categories in which LLMs can help automate learning activities: profiling/labeling, detection, grading, teaching support, prediction, knowledge representation, feedback, content generation, and recommendation. From the perspective of educators, there are no objections related to technology acceptance (Chocarro et al., 2023). Nevertheless, there are significant unresolved ethical issues, such as lack of reproducibility and transparency, and insufficient privacy measures (Yan et al., 2023).

Collectively, the literature review shows that AIEd has emerged as a serious cornerstone of HE – although it is considered still as a field to be developed (Charbonneau-Gowdy et al., 2023) – and LLMs in particular may have significant impact on teaching. Accordingly, the subsequent investigation on the potential of LLM-based ChatGPT is an important measurement to align the HE relevant regulations.

# 3 Materials and Methods

The study is based on a bachelor thesis, written by the first author from February to June 2023 and supervised by the two other authors in common weekly meetings. The objective of the bachelor thesis was to identify potentials and challenges of ChatGPT for HE based on environmental engineering as reference. In a second meta-level assignment part, the use of ChatGPT for the development of this bachelor thesis should be documented and analyzed. Neither ChatGPT nor other AI-assisted tools were used to write this article.

## 3.1    Bachelor Thesis

Methodically, the bachelor thesis was divided into four sections.

1. **Literature Review.** The theoretical foundation was built on a literature research, which allowed to outline the importance for LLM in teaching (see section 2).
2. **Practice Study.** A practical study, i.e., the application of ChatGPT to a final exam, was used to assess the performance of ChatGPT.
3. **Assessment Model.** The third step was to develop a simplified evaluation model for the suitability of an LLM to support the learning performance of a teaching-learning activity. The goal of such an evaluation model was to identify teaching-learning activities that particularly benefit from the availability of LLMs.
4. **Autoethnography.** Autoethnography was to be used to document all decisions to use ChatGPT as well as all ChatGPT chat history created during the development of the bachelor thesis.

## 3.2    Results

**Practice Study.** A final exam of the course *Urban Engineering Water* consisting of open questions was answered through ChatGPT. The answers of the one-hour exam with 18 questions were graded by the regular instructor of the course. In a first iteration, 51% of the points were obtained, based on a rigorous assessment, leading to a passed exam. This was followed by a second iteration in which ChatGPT was instructed to pay particular attention to those three books on which the course was based. The responses scored 61% of the points, using the same assessment scheme as before. The assessing instructor, who is to be described as quite demanding regarding digital tools in HE, was appreciatively surprised by ChatGPT's performance. Thus, it could be shown that ChatGPT might be used to pass a final exam. Admittedly, this study is slightly theoretical, as the large-scale use of ChatGPT required in a proctored exam is challenging to achieve.

**Assessment Model.** Purpose of the multi-criteria assessment model – developed on the theoretical foundations of multicriteria decision analysis (MCDA) (Belton & Stewart, 2002; Eisenführ et al., 2010; Söbke & Lück, 2022). – was to provide a measurement of the suitability of ChatGPT to support teaching formats. The objective was to provide a simple multi-criteria assessment model that could roughly assess the usefulness of ChatGPT for the learning effectiveness of a teaching format. Based on a literature review, the model was discussed and specified by the authors in an interactive session. It is vital to stress that this model is only a measure of the extent to which the learning functionality of a teaching format is enhanced by ChatGPT, but not - as described above - what effect ChatGPT has on the assessment functionality of a teaching format.

We choose interactivity (or: What is the regular level of interactivity of the teaching format?) as a criterion because interactivity is important for high learning effectiveness (Hattie, 2008) and is promoted by discussion and reflection of the learners (Dolfsma, 2002). According to Van Laer and Elen (2017), interactivity may be seen as a foundation for so-called blended learning environments, which drives the design of effective teaching formats (Slavin, 1996). ChatGPT is considered to provide interactivity into teaching formats from the students' perspective and thus be conducive to learning outcomes.

*Feedback* (or: What is the regular level of feedback given to the student within the teaching format?) was chosen as a criterion because formative feedback, in addition to individualizing teaching formats (Hattie, 2008), also supports student learning within blended learning environments (Van Laer & Elen, 2017), which according to Trigwell and Prosser (2004) impacts learning outcomes. Again, we see ChatGPT as a tool that may support the learning process through feedback.

We applied values for these two criteria to prevalent teaching formats and determined a suitability of ChatGPT to support the learning for these teaching formats (Table 1): Methodologically, the two criteria were each rated in three parts by the author team in an interactive session as high (3), medium (2), and low (1). For the suitability of ChatGPT, the mean of both criteria was used, rounded, and inverted. For example, for a thesis, interactivity is to be considered low (since individual work). The feedback is likewise to be estimated as low, since also here predominantly individual work takes place and actual feedback has to be always requested by the learner actively, for example by consultations with advisers. On average, this results in a value of 1, which inverts –because ChatGPT might remedy this shortcoming – to a value of 3 (high) for the suitability of ChatGPT to support learning during developing a thesis. We argue that this rating reflects reality well: In our assessment, ChatGPT can be seen as very helpful for learning during a thesis due to features such as structuring and summarizing text. In this assessment, we assume that the results of ChatGPT are critically reflected by the learners and thus foster learning.

**Table 1.** Suitability of prevalent teaching formats to foster learning based on ChatGPT support

| Teaching-Format | Interactivity | Feedback | Suitability of ChatGPT |
|---|---|---|---|
| **Thesis** | low | low | high |
| **Lecture** | low | medium | high |
| **Mandatory Homework** | medium | low | high |
| **Tutorial** | medium | medium | medium |
| **Excursion** | medium | medium | medium |
| **Practice** | high | high | low |
| **Groupwork** | high | high | low |
| **Consultation** | high | high | low |

**Autoethnography.** All considerations regarding the use of ChatGPT were documented in terms of autoethnographic logs (Belbase et al., 2008; Ellis et al., 2010, 2011; Raab, 2015) regardless of whether ChatGPT finally was used or not. Similarly, all chat histories were retained with ChatGPT. In total, five log entries (ca. 5000 words total) as well as 15 chat histories (ca.1200 words total) were available for the following analysis. In addition, the log entries were supplemented by a-posteriori reflections of the first author. Based on a rough estimate of the first author, the final bachelor thesis is based

1. textually on 1 % texts generated from ChatGPT 1% (measure: ratio of word-for-word adoption of ChatGPT (compared to complete text),
2. structurally on 15 % items suggested by ChatGPT 15% (measure: structuring elements, such as headings and lists)
3. ideationally on 10 % ideas suggested by ChatGPT 10%. (measure: ideas, which are explicated in the thesis)

The autoethnographic data were aligned with the nine recommendations for learners – as the autoethnographic data has been collected by a learner – by Gimpel et al. (2023). Gimpel et al. (2023) is a guideline  for instructors and learners for handling of ChatGPT and other LLMs, which was published in March 2023 as a result of the collaboration of a panel of experts from HE. These nine recommendations were chosen by us because they foster constructive use of ChatGPT and therefore, in our view, might be a benchmark for ChatGPT use.

## 4    Autoethnographic Experiences

For the following, the autoethnographic data are presented according to the recommendations of Gimpel et al. (2023). For each recommendation, a description of the recommendation itself is given first, followed by selected autoethnographic experiences made during the writing of the bachelor thesis, and finally a summary, striving to abstract the ChatGPT experience into generally applicable findings. When spoken of here in the first-person singular, this is written from the perspective of the first author, who is the author of the autoethnographic data.

## 4.1    Respect the Law and Examination Regulations

**Recommendation.** This recommendation indicates that learners have to observe applicable laws and regulations and, where appropriate, label AI-generated content including the information provided.

**Experiences.** Essential for the use of ChatGPT in the development of the bachelor thesis was the orientation towards applicable guidelines for the handling of artificially generated content. Here, the focus was on a self-responsible use of ChatGPT under consideration of prevailing regulation frameworks, such as the examination regulations of the HE institution. These, as well as further modalities concerning the development of theses had to be found out and implemented during the development of the thesis.

In addition, given the technical capabilities and limitations of ChatGPT, such as in generating outcomes, plausibility had to be checked. Since I have already successfully taken all courses of my study program at the time of the release of ChatGPT in November 2022, this was my first application of ChatGPT. My hope was that ChatGPT would help me to work more effectively in a more structured way and that ChatGPT could take handle some chore work, e.g., during the literature research and the writing. A review of the study regulations revealed that they did not contain any rules on artificially generated content. Also, no information in this regard was given in the preparatory course Scientific Work. Accordingly, I assumed that AI generated outcomes could be used when marked as third-party content. In addition, I was explicitly advised by the other supervising authors to use ChatGPT for writing the bachelor thesis and to document the results of the interactions between me and the AI. This request was also documented in the assignment of the bachelor thesis.

**Summary.** There was an understanding of relevant regulations, but it was apparent that the HE institutions regulatory framework did not fully account for the use of AI, including ChatGPT. By documenting the use of ChatGPT, its involvement was made transparent. However, it is important to note that this use of ChatGPT for research needs to be considered an exception, and therefore these findings may not be applicable to all theses. Moving forward, it may be beneficial to make documenting the use of ChatGPT, such as by including all chat histories, standard practice.

## 4.2    Reflect on Your Learning Goals

**Recommendation.** This recommendation indicates that interactions with ChatGPT need be done in a structured way. First, the learning goals must be clear. The learning goals drive the information that ChatGPT should provide to support the learning goals. The information requested result in chat commands (prompts) to ChatGPT. Finally, the outcomes generated has to be checked for plausibility. In summary, the achievement of domain-specific learning goals requires the mastery of a variety of ChatGPT related-meta skills, such as digital literacy and critical thinking.

**Experiences.** When developing the bachelor thesis, the learning goals are rather to be considered as information goals. Primarily, it is important to find information that can be processed in the bachelor thesis, only secondarily, it is also important to learn in terms of internalizing new knowledge. Accordingly, the main goal was the completion of the bachelor thesis. Nevertheless, knowledge goals were also defined, such as a literature review, the development of an assessment model, and the conduct of a study. In addition, autoethnographic experiences about incorporating ChatGPT in developing the bachelor thesis were to be documented. Here, details such as the structure and handling of interactions with ChatGPT were developed during the writing process.

I swiftly developed the following process structure:

a. **Awareness of learning objectives:** I had to be aware of my learning goals for supporting ChatGPT with necessary information. These include the context (writing a bachelor thesis), the prompts used (e.g., generating a chapter outline), and the expected outcome of the interaction (e.g., an outline).

b. **Development of prompts:** Based on learning goals and the expected outcomes, such as an outline or the generation of ideas via brainstorming, prompts had to be developed first. These were to guide ChatGPT to specifically generate the expected outcomes (see Chat 1).

c. **Content validation:** Afterwards, the outcomes had to be checked if they provide the expected information. In addition, a technical review (validation) had to be performed. On the one hand, the validation was based on my knowledge (self-expertise) and an assessment of the applicability of the outcome based on external, easily accessible resources, such as Wikipedia. On the other hand, validation often also required in-depth research, such as searching for other sources, e.g., articles or books).

d. **Reflection of the learning goals:** Reflecting on the outcomes again raised my awareness of the learning goals. This revealed whether valid outcomes were generated by ChatGPT. These reflected new learned knowledge, whereas unsuitable content was discarded. If necessary, consideration was given to adjusting the context, the prompts used, or even the interaction goal. Content validation provided learning in both positive and negative cases.

For example, I used ChatGPT to request 10 references about the use of AI in HE (see Chat1). Here, I pursued the goal of matching the previous results of my research and generating suggestions - keywords - for further literature research. By instructing ChatGPT to refer to literature reviews, technical reports, and meta-studies, I specified my prompts for obtaining scientifically relevant references.

**Chat 1.** Literature research

| | |
|---|---|
| **NS** | Find 10 references that address the use of AI in higher education. Preferably use literature reviews, technical reports, and meta-studies. (Chat1_Prompt1_Topic2[1]) |
| **GPT** | Bower, M. (2019). Decoding the Discourse: A Systematic Review of Machine Learning Applications in Education Research. Review of Educational Research, 89(5), 785-827.<br>Kovanović, V., Gasevic, D., Joksimović, S., & Hatala, M. (2018). Applying machine learning in education: Toward personalized learning. In Handbook of Learning Analytics (pp. 159-182). Society for Learning Analytics Research.<br>[…] |

ChatGPT reliably returned 10 references. I started the validation process with author verification via Google Scholar author search for confirming the existence of the named authors and their subject specialization. Subsequently, I checked the named journals and conferences for their existence, discipline, and year of publication. Article titles were searched via Google Scholar and Scopus, both in full and in sections. Finally, I validated discoverable references based on abstracts and conclusions. This revealed that 6 out of 10 references depicted literature reviews, technical reports, and meta studies. However, it turned out that about half of the authors existed but were not in the relevant research areas (AI, IT, computer science, education). In addition, only one of the 10 references was immediately available under the stated title and authors, and one was further available under a slightly different title. Upon review, 9 of 10 references proved to be implausible. As a result, I weighed the benefits of ChatGPT in the further literature search and decided not to use ChatGPT for this purpose anymore due to mostly invalid outcomes.

---

[1] Chat1_Prompt1_Topic2 references the ChatGPT history created in this study

**Summary.** The use of ChatGPT requires a clear structure that emanates from the learning objectives. Further, the development of suitable prompts and the validation of the outcomes are of particular importance. Using ChatGPT in the literature review made evident that ChatGPT did not meet the achievement of the learning objective – provision of relevant references – despite the iterative development of prompts. However, ChatGPT was beneficial in the generation of further keywords for the research process.

## 4.3 Use ChatGPT as a Writing Partner

**Recommendation**. This recommendation indicates that ChatGPT may serve as a writing partner for various writing-related activities, such as brainstorming, text structuring, and writing. In doing so, the acquisition of digital competencies is necessary. These competencies include the purposeful development of prompts as well as the validation of the results. Further, users need awareness about the limitations of ChatGPT.

**Experiences.** When working with ChatGPT as a writing partner, brainstorming interactions were mainly used. From these, about 10% of the bachelor thesis resulted from suggestions for text structuring and the evaluation of text sections. Here, I interacted with ChatGPT not having any concrete expectations about the usefulness of the generated structure. This reservedness was grounded in my experience with ChatGPT during the literature research (see chapter 4.2), where I became aware of the outcomes generally to be questioned critically. Nevertheless, I hoped to be able of adopting the generated chapter structure for my thesis at least in parts. Resulting I prompted ChatGPT for structuring of a chapter about the definition and functionality of ChatGPT. To get an answer as clear and simple as possible, I limited ChatGPT to the output of bullet points (Chat 2). I also did this to adapt ChatGPT to my previous brainstorming outcomes. The focus was to ensure that I had freedom in the further development of the chapter (selection of topics, prioritization, form of presentation) and had not simply to adapt an outline that already seemed logical in itself without critical validation and my suggestions.

**Chat 2.** Chapter structuring

| | |
|---|---|
| **NS** | How could a chapter of my bachelor thesis be structured, which deals with the definition and functionality of ChatGPT? Generate 5 bullet points without further explanations. (Chat2_Prompt1_Topic1) |
| **GPT** | Of course, here are five key points for structuring the chapter of your bachelor thesis on "Definition and functionality of ChatGPT": |
| | 1. introduction to ChatGPT |
| | 2. basics of the GPT architecture |
| | 3. components of ChatGPT |
| | 4. functionalities in practice |
| | 5. challenges and further development |

ChatGPT generated the expected structure and suggested to include a thematic introduction as well as basics of the GPT architecture (see Chat 2). I had not considered a more detailed description of the GPT architecture until then. Nevertheless, I found the idea suitable and worked out an appropriate structure. Here I specifically explained the concept behind the GPT architecture. Furthermore, points 3,4 and 5 coincided with my outline of the chapter about the definition and functionality of ChatGPT. Accordingly, I began to include these as subchapters in my thesis as well.

In developing text and chapter structures, the focus was on reviewing already developed texts. For example, I asked ChatGPT whether the given text section on ChatGPT's capabilities and limitations addressed all relevant aspects. In relation to this, I had already read a large number of studies, some of which relied on very different data. Therefore, I decided to take an approach from other studies on ChatGPT and survey ChatGPT about themselves to get accurate information about ChatGPT-specific issues. In this way, I hoped to get the confirmation of my texts by ChatGPT. Thus, I pursued getting an overview and not a lengthy output of ChatGPT. Accordingly, I limited ChatGPT to the generation of 5 key points.

**Chat 3.** Evaluation of the argumentation structure

| **NS** | Check the following text for the relevance of its argument structure. Are all relevant aspects taken into account? Generate 5 key points without further explanation. [Text] (Chat3_Prompt1_Topic4) |
|---|---|

| **GPT** | The text takes into account all the essential and relevant aspects in their argumentative structure: |
|---|---|
| | 1. emphasis on strengths and limitations in clear alternation. |
| | 2. mentioning versatility and natural language processing as strengths. |
| | 3. highlighting limited world knowledge and lack of judgment as limitations. |
| | 4. inclusion of scalability and fast response times as positive features. |
| | 5. consideration of lack of empathy and susceptibility to misinformation as weaknesses. |

According to the evaluation of ChatGPT (see Chat 3), my argumentation considers all key points, contrary to my assumption. In addition, 5 aspects highlighted the relevance of my reasoning. Hereby, by stating some capabilities and limitations of ChatGPT, my confidence was raised that ChatGPT seems to be familiar with the subject matter and that the assessment generated was conclusive. At this point I realized that regardless of whether human or machine, a conclusive and above all positive assessment of one's own performance generated confidence for the further work process. As a result, I perceived those certain interactions with ChatGPT triggered an emotional process. Subsequently, I trusted the expertise of ChatGPT and finished the work on the chapter with a clear conscience.

In addition, ChatGPT helped with the formal review of already written sections. The goals of the review were, among others, the writing style, the technical terms, as well as the elimination of filler words. Accordingly, I asked ChatGPT to evaluate a section of my conclusion of the bachelor thesis regarding an objective writing style and to shorten filler words. I refrained from a full evaluation because it bothered me to submit several pages to ChatGPT compared to a smaller section. In addition, I could not refer to appropriate guidelines and experience reports, which guided such a processing of data. In this regard, I hoped that ChatGPT would issue an improved version of my text section in a more pointed manner and using technical terms. Accordingly, I asked to replace colloquial expressions with technical terms and phrases in more academic style. The generated text served as a template for me to make selective changes at my own discretion. This was done because I was convinced that my expertise would allow me to make more qualitative changes manually than if I had simply applied the ChatGPT outcomes. In addition, I assumed that I would be able to make statements about the quality of the entire conclusion on the basis of the template.

**Chat 4.** Writing style evaluation

| **NS** | Does the conclusion section of my bachelor thesis use an objective writing style? Replaced colloquial expressions with technical terms: [[...] As a result, limitations that currently inhibit the qualitative use of AI tools as aids in higher education can be replaced by application-specific functions in the foreseeable future]. (Chat4_Prompt3_Topic3) |
|---|---|

| **GPT** | The paragraph from the conclusion of your bachelor's thesis already uses a relatively objective writing style. Here is a version with some colloquial expressions replaced by technical terms: |
|---|---|

[[...] As a result, currently existing limitations that currently hinder a qualitative integration of AI tools into university teaching could be substituted by application-specific functionalities in the medium term].

From ChatGPT's response, I inferred that I could still make changes regarding my writing style to improve the quality of my conclusion. I was also curious on the changes ChatGPT had made and whether I might learn any new technical terms. After comparing both texts, I integrated a few new technical terms into my work at my discretion, i.e., I could adapt the basic vocabulary of my work based on ChatGPT. In doing so it was important to me that I personally identified with the text and did not simply adopt generated content. As a result, I looked at the generated section of ChatGPT (see Chat 4) very critically. While I trusted ChatGPT's expertise regarding writing style, I was not always sure of the changes made by ChatGPT. Unlike ChatGPT, I had finally developed a context and made thematic prioritizations based on experience from many hours of work and over 120 pages written. On the other hand, ChatGPT generated a new version based on a small section of text, which in some places reflected the meaning of the content but changed prioritizations and context.

**Summary.** In conclusion, ChatGPT could be used as a versatile writing partner. Further, the relevance of building digital literacy skills in prompt development became apparent. Particularly through specifications regarding the length, structure, and content of sections to be generated, individual adjustments could be made. These facilitated the work process with ChatGPT. In addition, obviously that awareness of ChatGPT's limitations was conducive to a critical reflection of the results: The limitations sharpened the view of which responses of ChatGPT in particular are to be reflected, such as in Chat 4. It was important to interact with ChatGPT without the expectation of actual perfect outcomes to avoid demotivating frustration. As a result, confidence in the generated content of ChatGPT could be built up, which on the one hand accelerated the work process and on the other hand gave more security.

### 4.4 Use ChatGPT as a Learning Partner

**Recommendation.** This recommendation indicates that ChatGPT may be used systematically as a learning partner. This possible use of ChatGPT has already been addressed in the evaluation model of the bachelor thesis. On the one hand, ChatGPT acts as a tutor, which moderates new learning content for the user. Thereby, adaptations of the learning content may be made according to the user's cognitive abilities, competencies and needs. On the other hand, the user can access feedback via ChatGPT independent of time and place in order to actively support the learning process even outside of formal learning activities.

**Experiences.** During my work I used ChatGPT for brainstorming. I hoped to identify relevant aspects of my work with less mental effort. Again, I had no expectations regarding the usefulness of ChatGPT's outcomes for my work. Also, comparisons with developed ideas were mostly missing. However, I needed to validate the outcomes generated by ChatGPT, an activity that led to a learning process. Thus, I challenged ChatGPT to demonstrate a possible integration of AI in HE exclusively based on 3 to 5 aspects (Chat 5). In this, I limited ChatGPT. After all, I wanted to stimulate my thoughts with new suggestions, but without replacing it with prefabricated content or influencing it too much. Therefore, it was important for me to decide about the further procedure based on my thinking. Thereby I felt mostly strongly limited by already existing outcomes of ChatGPT, so that on my part no full-fledged brainstorming could take place.

**Chat 5:** Brainstorming

**NS** Show a possible integration process of AI in higher education teaching based on 3 to 5 aspects. Answer only in bullet points without further explanation. (Chat5_Prompt1_Topic6)

**GPT** Of course, here are 5 aspects that show a possible integration process of AI in higher education teaching:

1. needs analysis and goal setting

2. selection of appropriate technologies and tools

3. pilot phase and training

4. successive integration and further development

5. monitoring, evaluation and adaptation

As a result, I realized that ChatGPT had taken away part of the visualization of my thought process. This made brainstorming easier, but without influencing my thoughts too much by existing outcomes and taking away the decision about the further work from me. On the other hand, I was surprised that ChatGPT had mapped quite a complex process using such simple cues. On this occasion, I was impressed by the power of ChatGPT. Concluding, based on the 5 generated aspects (Chat 5), I worked out an integration process of AI into HE, in which I combined 2 of the mentioned aspects into one each. In conclusion, I was able to develop my representation of the integration, which was based on 3 aspects (1. perception, 2. enrichment, 3. integration).

Further, the interactions with ChatGPT as a learning partner were also based on the preparation of the colloquium of the bachelor thesis. Thus, I presented a summary of my bachelor thesis to ChatGPT, which I had evaluated once again regarding the thematic relevance for a technically foreign audience, since my two supervisors were interested in the work not as AI experts, but as instructors who are interested to integrate ChatGPT into their courses.

Thereby I enjoyed the easy accessibility of ChatGPT and interacted location-independent during a train ride with ChatGPT. In doing so, I requested an assessment from ChatGPT in a short time, which allowed me to rank the quality of my summary. Rather, the request followed my interest in what such an assessment of ChatGPT looked like. In addition, I hoped to gain confidence about my work from ChatGPT's feedback.

**Chat 6.** Evaluation of thematic relevance

**NS** Rate the following summary of my bachelor's thesis in terms of relevance and comprehensibility to an audience outside my field in 5 bullet points. (Chat6_Prompt1_Topic2)

**GPT** Here is how to evaluate your bachelor thesis abstract for relevance and comprehensibility:

Relevance: The summary addresses important aspects of AI research in education and clarifies its future implications. ⭐⭐⭐⭐⭐

Comprehensibility: the language is clear and understandable, despite the technical terminology. Readers should be able to grasp the main points. ⭐⭐⭐⭐

Topic Complexity: The summary adequately addresses AI developments and educational use without going into too much detail. ⭐⭐⭐⭐

Clarity of conclusion: the conclusion emphasizes the need for HE responses to AI use, but the link to "disruptive impact" could be clearer. ⭐⭐⭐

Overall Impression: the summary is relevant, understandable, and addresses important future trends, but could be a bit more specific in the conclusion. ⭐⭐⭐⭐

Overall rating: ⭐⭐⭐⭐ (4 out of 5 stars)

ChatGPT pointed out the strengths of my summary. I was surprised that ChatGPT had generated unprompted stars that visualized the rating (see Chat 6). The stars delighted me similar to good grades in high school. In addition, ChatGPT made it possible to prepare my content independent of location and time.

**Summary.** ChatGPT proved to be usable as learning partner by low-threshold provision of feedback independent of time and place. Also, the contents of the bachelor thesis could be summarized and moderated. The ever-necessary validation of the ChatGPT outcomes was another base for learning. Furthermore, ChatGPT supported brainstorming by generating structured outcomes, e.g., by keywords. Additionally, ChatGPT's evaluation of texts contributes to learning experiences. In addition, surprising parts of the outcomes produced learning experiences.

## 4.5    Iterate and Converse with ChatGPT

**Recommendation.** This recommendation indicates that by iterating with ChatGPT, i.e., repeatedly adjusting the prompt after evaluation the results and re-entering the adjusted prompt, an optimization of outcomes may take place. Here, unspecific answers of ChatGPT can be specified by improved prompts. In addition, through reinforcement learning and learning analytics methods, ChatGPT may adapt future content generation to the user's "preferences".

**Experiences.** By iterating with ChatGPT, outcomes may be adapted to personal preferences and requirements - e.**g.,** the scientific writing style of a bachelor thesis. On the one hand, ChatGPT's outcomes may be adapted by concretizing the prompt. Thus, besides the writing styles available to ChatGPT, grammatical preferences or forms of presentation preferred by the user, such as the answering of questions in bullet points, can be specified. The idea for this came from a YouTube video from 2023 (Digitale Profis, 2023), which I had picked out in preparation for working with ChatGPT. This video suggested amongst others enriching the prompts for ChatGPT with as much information as possible, such as context, length and the writing style to be used. As a result, I had the idea to ask ChatGPT about writing styles available to provide my prompts with more precise information (Chat 7).

**Chat 7:** Inquiring about writing styles

**NS**    What writing styles are available to you when generating scientific texts? Write them down in a list. (Chat7_Prompt1_Topic3)

**GPT**    Of course, here is a list of writing styles I can use when generating scientific texts:

1. expository

2. analytical

3. argumentative

4. descriptive

5. comparative

6. critical

7. explorative

8. theoretical

9. empirical

10. interdisciplinary

11. literature review

12. synthesizing

Please note that I am able to adapt writing styles depending on the requirements and context of your academic paper.

ChatGPT generated a total of 12 writing styles, which I could use for a more precise design of prompts for my scientific work. In addition, a hint was generated, which pointed out the contextual use of different writing styles. Here I was surprised by the variety of styles available to ChatGPT.

As a result, I checked the effect of the writing styles in generating a text section about the perception of AI in German HE. In preparation, I checked the meaning of writing styles 3 (argumentative) and 7 (explorative). I assumed that the focus of the outcome would change dependent on the writing style. Specifying an argumentative style focused on presenting different views and rationales of the student and faculty respondents about the use of AI in HE. Divergently, an exploratory style focused on presenting an overview of the topic, which was not yet complete due to its topicality. The generated section resembled an introduction, which outlined the topic of AI in HE and selectively outlined opinions for and against its use. As a result, the specification of different writing styles influenced the context of the text section to be generated.

Here I noticed that about half of the writing styles designations do not constitute a writing style as such, but rather a research method. For example, writing styles 7, 8, and 9 are more likely to represent a research method. Nevertheless, this observation might indicate that ChatGPT is not only aware of various research methods but is also capable of adapting the outcomes to characteristics of research methods.

Describing another example, I noticed that in the course of longer interactions (10-20 questions), ChatGPT automatically adjusted the outcomes to my preferences. Thus, ChatGPT started to respond using bullet points. Further, content from past interactions was involved. For example, ChatGPT used an interaction about the potential uses of AI in HE to generate a list of curricular challenges of the AI integration process (Chat 8).

**Chat 8:** Conversation with ChatGPT

**NS** Name possible uses of AI in higher education teaching. (Chat8_Prompt1_Topic1)

**GPT** Artificial intelligence (AI) can be used in a variety of ways in higher education to enhance teaching and learning experiences. Here are some potential uses:

1. personalized learning support: AI can provide learners with individualized resources, exercises, and feedback based on their learning styles, abilities, and progress.

2. Automated grading: AI systems can automatically grade exams, assignments, and homework, recognizing specific criteria or patterns. This saves instructors time and enables faster feedback.

3. Adaptive learning platforms: Platforms that use AI can analyze learning progress and recommend to learners specifically what content they should work on next to promote optimal learning.

4. [...]

**NS** Create a list of curricular challenges of the integration process of AI in higher education. (Chat8_Prompt16_Topic2)

**GPT** Integrating artificial intelligence (AI) into higher education brings several challenges that need to be addressed in the curriculum. Here is a list of curricular challenges:

1. personalized learning support: developing algorithms to tailor learning content to individual needs requires an understanding of learning psychology and data analytics.

2. automated assessment: curriculum design to incorporate assessment methods suitable for AI support, e.g., multiple-choice questions, coding exercises, or simulation-based exams.

3. adaptive learning platforms: Integrate data analytics and machine learning into the curriculum to understand adaptive algorithms and teach their application on learning platforms.

4. [...]

ChatGPT listed 12 aspects with a short explanation in both interactions. Here ChatGPT created a list un-prompted for question 1, only for (Chat1_Prompt16_Thema2) this was explicitly requested. Thereby I had the impression of having seen the headings before in a similar form. Accordingly, I checked my chat history for potential influences, such as previous prompts, similar prompts, or duplication of prompts. I noticed headings that were identical to those of my last prompt. I included, that the answers to prompts on a comparable topic within the subject area of AI in HE is influenced by previous questions and their outcomes. Thus, ChatGPT generated matching curricular challenges based on the named uses of AI in HE. Here, the headings of the enumerations were congruent, only the explanations suggested belonging to a different prompt. At this point, I recalled my literature review. Some studies reported that increasing iteration and the use of contextual questions on ChatGPT trigger the reinforcement learning method. Here, ChatGPT can identify the user's requirements during the chat, and take them into account when generating future outcomes, drawing on previous instructions (OpenAI ChatGPT Generated Results & Ventayen, 2023; Rudolph et al., 2023; Susnjak, 2022).

These session-related adjustments captured my attention. Thus, Table 2 presents an exemplary chat history of successive 20 interactions. For each interaction, the prompts used are categorized according to their interaction goal (here: brainstorming, structuring, overview) and their topic. A remark describes the result of each interaction. In addition to the length and conciseness of the outcomes, the fulfillment of the interaction goal is also described. Here, in the case of Success, the interaction goal was achieved, the result was satisfactory. In the case of a Cancel (Table 2, Prompt# 10-12), the result was not satisfactory, so the interaction and further iterations were discontinued. In this case, the overall goal was to develop a thematic overview, which was generated in sufficient form regardless of whether the output was too long or too short, making further interactions obsolete. Overall, Table 2 provides an overview of the development of prompts. Such developments can be very time-consuming and lengthy, and the outcomes are not always successful.

**Table 2:** Prompt Iteration

| Prompt # | Category | Topic | Remark (Problem/ Success / Cancelling) |
|---|---|---|---|
| 1 | Brainstorming | Possible applications of AI in HE | Output too large |
| 2 | Brainstorming | Possible applications of AI in HE | Output too general |
| 3 | Brainstorming | Possible applications of AI in HE | Output too short |
| 4 | Brainstorming | Possible applications of AI in HE | Output too inaccurate |
| 5 | Brainstorming | Possible applications of AI in HE | Success |
| 6 | Overview | Challenges of AI in HE | Output too general |
| 7 | Overview | Challenges of AI in HE | Output too large |
| 8 | Overview | Challenges of AI in HE | Output too inaccurate |
| 9 | Overview | Challenges of AI in HE | Success |
| 10 | Overview | Perception of AI in HE | Output too large |
| 11 | Overview | Perception of AI in HE | Output too short |
| 12 | Overview | Perception of AI in HE | Cancel |
| 13 | Structuring | Integration process of AI in curricular | Output too inaccurate |
| 14 | Structuring | Integration process of AI in curricular | Output too large |
| 15 | Structuring | Integration process of AI in curricular | Success |
| 16 | Brainstorming | Curricular challenges of AI integration in HE | Output too short |
| 17 | Brainstorming | Curricular challenges of AI integration in HE | Output too general |
| 18 | Brainstorming | Curricular challenges of AI integration in HE | Output too inaccurate |
| 19 | Brainstorming | Curricular challenges of AI integration in HE | Output too large |

| 20 | Brainstorming | Curricular challenges of AI integration in HE | Success |

**Summary.** Entire chat histories based on prompt iterations with ChatGPT demonstrated that the definition of the work goal and the structuring of prompts, is highly relevant for the outcomes. Nevertheless, it has been shown that iterations have allowed the outcomes to be aligned with the work goal. For example, prompt characteristics (see Chat 7) may be requested, which specify the design options of prompts. As a by-product of prompt iterations, the user's requirements (e.g., form of presentation, text composition, writing style) are incorporated over time as preferences in the outcomes of ChatGPT in an automated manner, and the need for precisions is eliminated.

### 4.6    Summarize Learning Material with ChatGPT

**Recommendation.** This recommendation indicates that ChatGPT may be used to summarize learning content in a more concise manner. Text sections can be reduced to their essential core in a short time. Thus, a targeted adaptation of the learning content to the individual needs of the user supports learning.

**Experiences.** The contents of the bachelor thesis are regarded herein as learning material. I used ChatGPT for the extraction of contents of scientific articles. This was beneficial especially during the literature review. Actually, I provided selected parts of an article, such as the abstract, to ChatGPT for generating a summary. Thereby, I hoped of being enabled to judge about the suitability of a study regarding aspects, such as relevance, method, topic, and findings in a short time. Firstly, this was done based on my expertise of the abstract and the summary generated by ChatGPT. I also provided ChatGPT with additional text sections besides the abstract to summarize long studies and to clarify passages that were technically complex for me. Hereby I limited ChatGPT to the generation of 5 key points. Thus, I wanted to be presented the contents of the study clearly. For example, I created the prompts to a systematic review of Zawacki-Richter et al. (2019), to summarize the abstract in 5 bullet points. In this case, there were no comprehension problems, rather I wanted to check whether this way of reviewing references was meaningful for me.

**Chat 9.** Study Summarization

| NS | Summarize the following abstract of a Systematic Review on the uses of AI in higher education. Limit yourself to 5 key points: [Abstract] (Chat9_Prompt4_Topic2) |
|---|---|
| GPT | • AI in Education (AIEd) is an emerging field in educational technology.<br>• Educators struggle to effectively utilize AI for pedagogical benefits in higher education.<br>• A systematic review of 146 articles from 2007 to 2018 reveals AIEd trends.<br>• Dominant disciplines are Computer Science and STEM, quantitative methods prevail.<br>• AIEd applications: profiling, assessment, adaptivity, intelligent tutoring; lacks critical reflection, pedagogical integration, and ethical exploration. |

ChatGPT generated as requested a summary from 5 key points. Additionally, to my assessment of the abstract, I was able to use the key points for developing an overview of the article for the bachelor thesis. Here I was impressed by the effectiveness of ChatGPT to summarize text sections in only a few key points. As a result, ChatGPT supported me to develop a meaningful bibliography using a reference manager software (Zotero). I was able to annotate the articles with outcomes from ChatGPT. Thereby, I annotated each article with notes regarding their main statements. This greatly facilitated my writing, as I created folders in the reference manager software for each chapter of my bachelor's thesis. Afterwards, I used the notes, restructured and rephrased them. Finally, the references used in the process were inserted.

Apart from the literature research, ChatGPT helped me to summarize the content of the bachelor thesis during writing. I used ChatGPT for creating summaries of my texts for including all relevant aspects in the conclusion of a chapter. This approach also gave me an advantage in the later selection of contents for my colloquium, in which I mostly referred to my conclusions of single chapters of my thesis. In doing so, I instructed ChatGPT to summarize conclusions I had gathered in bullet point form. This approach was used for the literature review, feasibility study, evaluation model, and autoethnography, which were the most effort-consuming chapters. ChatGPT helped me not missing any aspect due to a careless mistake.

**Chat 10:** Conclusion Summarization

| **NS** | Summarize the following conclusion about the prevailing views of AI in higher education. Limit yourself to the output of 5 bullet points. (Chat10_Prompt3_Topic5) |
|---|---|
| **GPT** | • Acceptance & utility value of AI in higher education available<br>• User experience & skills shape user awareness<br>• Importance of AIEd at German universities increased<br>• Quality over abundance in AI applications<br>• Consistent teaching views for acceptance & collaboration |

ChatGPT generated as requested a summary in 5 bullet points. These formed the relevant summary of the chapter after my review of my conclusion. As a result, I was able to use ChatGPT in the development of the presentation slides for my colloquium to summarize the content of 3 chapters in a few bullet points. Here, the focus was on adapting the contents of my bachelor thesis with the help of ChatGPT in order to then organize them for the preparation of the colloquium.

**Summary.** The chats with ChatGPT demonstrated that learning content may be efficiently summarized in a short time. ChatGPT is able to prepare learning content according to the user's needs and adapt it for further work. Also, by employing suitable prompts, learning content can be adapted to the individual work process. However, it appears necessary to organize the contents of each chapter outside of ChatGPT for summarizing them. As a result, ChatGPT is merely a tool that does not take over the preparation of learning content but requires skills of the user to prepare the content meaningful.

### 4.7 Boost Coding with ChatGPT

**Recommendation**. This recommendation indicates that ChatGPT may also be used for programming. Here, assistance may be given, programming templates may be recommended and also errors might be identified and fixed.

**Experiences.** During my research on ChatGPT I came across its capabilities for generating program code and wanted to explore them myself - even though it was not part of my bachelor thesis. Here I remembered a lecture in a computer science course about the function of while loops in the programming language Java. Thus, I refreshed my knowledge about while loops and asked ChatGPT for an explanation and a small programming example. Here I was especially curious about the comprehensibility of the code example, because in the course Java programming appeared complicated to me.

**Chat 11.** Learning code by interest

> **NS** What is a while loop in Java? Also, give me a short code example as an illustration. (Chat11_Prompt1_Topic1)

> **GPT** A while loop in Java is a control structure that allows you to repeatedly execute a block of code as long as a specified condition evaluates to true. The loop keeps executing as long as the condition remains true, and it stops when the condition becomes false.
>
> Here's a short code example of a while loop in Java:

```java
public class WhileLoopExample {
    public static void main(String[] args) {
        int count = 0;

        // This while loop will execute as long as the count is less than 5
        while (count < 5) {
            System.out.println("Count: " + count);
            count++; // Increment the count by 1 in each iteration
        }

        System.out.println("Loop finished");
    }
}
```

In this example, the while loop will iterate five times because it starts with count being 0 and increments it by 1 in each iteration until it reaches 5. The output will be:

```vbnet
Count: 0
Count: 1
Count: 2
Count: 3
Count: 4
Loop finished
```

Please note that it is important to make sure that the condition in the while loop becomes false at some point so that the loop does not run indefinitely and form an infinite loop.

ChatGPT generated a small section about the definition and functionality of while loops (Chat 11). The code example was also understandable for me, although computer science has always been a minor subject for me. I was surprised that ChatGPT was able to prepare a topic that was previously not very clear for me in a comprehensible way.

Based on this good experience, I planned to spend more time with code programming in my spare time and to learn the basics of the programming language Python. Thus, I requested ChatGPT to generate a mini curriculum for me, which structured theoretical and practical learning content according to a pedagogical order. Conclusively, appropriate content, questions and programming test should be provided for me to acquire and check the knowledge level. Further, I stated that my theoretical and practical skills should be checked through tests to evaluate my knowledge level.

**Chat 12:** Learning Code by ChatGPT

**NS** Create a curriculum that introduces me to the theoretical and practical basis of programming in Python. Include in it questions and programming tests to check my level of knowledge. (Chat12_Prompt3_Topic4)

**GPT** Absolutely, here is a structured curriculum to teach you the theoretical and practical basics of programming in Python. The plan covers several stages and includes lessons, practical exercises and self-tests:

**Level 1: Introduction to the Basics**

**Lesson 1: Introduction to Programming**
- What is programming?
- Different types of programming languages
- The programming process

**Lesson 2: Introduction to Python**
- History of Python
- Installing Python
- The interactive Python shell

**Lesson 3: First steps with Python**
- Output of text (print function)
- Variables and data types (e.g. int, float, str)
- Comments

**Practical Exercise 1: Basic Outputs and Variables**
- **Write a program that outputs your name and age.**
- **Calculate the average of three numbers.**

**Self-test 1: Basics**
- **What is the purpose of the print function?**
- **What data types do you know, and how are they declared in Python?**
- **How do you comment code in Python?**

**Level 2: Control structures and functions**

**[...]**

**Final Project Proposal:**
- **Develop a simple task management application with a user interface that allows users to add, edit, and mark tasks.**

ChatGPT generated a total of 12 lessons with three suggestions for learning materials each. In addition, a total of 4 self-tests and 4 practical exercises were generated, for assessing the knowledge after the completion of every 3rd lesson. As a result, I tried the learning contents of the first lesson for judging the applicability of the curriculum generated.

So, as part of the first lesson, I asked ChatGPT about the meaning of the term programming. ChatGPT provided an overview of the definition, functionality and meaning of programming in computer science. Since here for me, educated as an environmental engineer, questions remained open, I iterated driven by curiosity about the impact of programming in industry during the last 20 years and which areas could be influenced by it substantially. As a result, ChatGPT generated a text more than one screen long, which referred to the increasing digitalization process of the industry - Industry 4.0 - and presented new professional fields opened up by it. In conclusion, there was a longer chat with ChatGPT in which I worked out the topic according to own interest on the basis of the generated curriculum and let ChatGPT prepare and structure the content and visualize it by generating code examples (see Chat 11).

Although I had no competencies in programming with Python, I considered the generated curriculum as coherent. All the essential content for acquiring a basic knowledge of programming with Python was addressed in my perspective. In addition, I was excited by ChatGPT's suggestion of tackling a larger programming project as a final exercise. Overall, based on my experience with the first lesson, I could imagine continuing to learn Python following the curriculum generated by ChatGPT. Further, I was impressed by the option to execute lessons multiple times if required and by the option to ask for feedback (see section 4.4. Use ChatGPT as a learning partner). I was satisfied that I could validate ChatGPT's outcomes using a python interpreter without much mental effort.

**Summary.** ChatGPT offers various options in the field of programming. On the one hand, ChatGPT can be used as a tutor to teach learner-oriented competencies in in the field of programming. Thereby the user has many options, e.g., guided by a curriculum or interactively as a learning partner, to actively design his learning process with ChatGPT. In addition, ChatGPT might be used for troubleshooting: ChatGPT can generate targeted suggestions that allow the user to go his own creative ways at any time when programming, which are not restricted by the specifications of a curriculum.

### 4.8 Beware of Risks When Using ChatGPT

**Recommendation.** This recommendation points out that interaction with ChatGPT is associated with risks. In addition to false statements, ethically questionable, e.g., discriminatory, content can also be generated. As a result, there is a need for awareness of applicable guidelines in the use of ChatGPT. Furthermore, in the context of preparing scientific papers, artificially generated content must be clearly declared as such and checked for plausibility.

**Experiences.** Crucial was an awareness of guidelines for handling artificially generated content during thesis writing (see section 4.1). The documentation of interactions can contribute to a transparent use of ChatGPT. For example, chat logs might be provided via attachments, for example. Nevertheless, the use of AI - especially LLM like ChatGPT - is not yet anchored in the regulatory framework of the HEI.

Especially when working with ChatGPT as a research tool (see section 4.2), the focus was on applicable guidelines of the HEI. However, due to the novelty of ChatGPT, no guidelines existed. As a result, I worked out a reference style for artificially generated content derived from the Harvard citation style to ensure the traceability, which is a criterion of scientific quality of the bachelor thesis. In addition, the validation of ChatGPT outcomes during the literature reviews revealed that almost no existing references were generated.

Awareness of the risks using ChatGPT also impacted the writing process (see section 4.3). It became apparent that knowledge about the limitations of ChatGPT could support the validation of ChatGPT's outcomes: Accordingly, I identified validated content susceptible to flaws with higher priority.

**Summary.** In conclusion, working with ChatGPT resulted in a reflective handling of ChatGPT's outcomes. Furthermore, based on the risk of erroneous statements and the lack of guidelines in dealing with artificially generated content in the preparation of theses, the interaction with ChatGPT was focused on selected areas of application, such as brainstorming, or structuring (see Table 2). Also, the creation of chat logs for the purpose of documenting the use of ChatGPT turned out to be useful especially for analyzing purposes of the bachelor thesis. For occasional use, the documentation of ChatGPT's user interface is beneficial.

### 4.9     Read This Checklist Before Using ChatGPT

**Recommendation.** This recommendation indicates that a checklist increases the quality of using ChatGPT. By prompt templates, prompt parameters and evaluation criteria and, the interaction with ChatGPT can be intensified. At the same time, the user's own responsibility for the implementation of given regulations is brought to the fore.

**Experiences.** The discussion of the recommendations in the previous sections shows, it turned out that almost all recommendations were relevant to the bachelor thesis. The previous study of the recommendations contributed to a more purposeful use of ChatGPT and accelerated the work process.

Thus, due to the non-existing guidelines at the HEI, the handling of artificially generated content was based on agreement with the supervisors of the bachelor thesis (see Section 4.1). Also, a referencing of ChatGPT outcomes in Harvard style was developed. In addition, chat logs captured interactions in a traceable manner. The focus was on the development of a working structure that, starting from the definition of the learning goal, provided orientation for the use of ChatGPT as a writing and learning partner (see Section 4.2).

**Summary.** Using ChatGPT when writing a bachelor thesis requires skills regarding the use of ChatGPT. The acquisition of such skills is facilitated by the use of this checklist.

## 5      Discussion

The implementation of ChatGPT in learning activities is an innovation that changes learning processes and has disruptive potential for certain learning activities, i.e., these learning activities become obsolete as the learning functionality of the activity is eliminated. On the other hand, ChatGPT also opens new opportunities, for example, it may be used for different activities that previously had to be accomplished without any assistance from a digital tool, such as brainstorming, structuring, and text revision. Accordingly, ChatGPT is also helpful in the development of longer texts, as the experience described here in writing a bachelor thesis demonstrates. Although it could be evidenced that ChatGPT is able to pass knowledge-oriented exams, ChatGPT currently is to be seen as a tool and not as a disruptive game changer when writing theses. Limitations, such as referencing non-existent sources, suggested that ChatGPT on its own is currently not capable of producing a bachelor's thesis of acceptable quality.

It should be noted that the use of ChatGPT calls for meta skills (Alotaibi & Alghamdi, 2022; Binkley et al., 2012). In particular, technical literacy skills specific to ChatGPT are required: Prompt engineering in particular, i.e., the ability to design questions and commands (prompts) to ChatGPT in a manner that ChatGPT responds with the expected outcomes. Furthermore, information literacy is demanded from users by the continuous monitoring of the output of ChatGPT for plausibility and also quality. While the ChatGPT specific technical literacy skills are rather to be learned additionally, the information literacy skills are considered to be beneficial for other activities as well.

When dealing with ChatGPT, especially prompt engineering was trained. Furthermore, ChatGPT had to be integrated into individual work processes, so that there was a change in the work process compared to other writing tasks already done. On the one hand, time could be saved by generating information and texts, on the other hand time had to be spent for a validation of the ChatGPT outcomes. Due to the validation, the work processes described here appeared to be rather more time-consuming. However, the lack of experience and the research context itself requiring documentation may also be regarded as time-consuming. Additional effort arises from the integration of ChatGPT outcomes and manually generated content as well.

In addition to the meta skills required by ChatGPT, there might occur also other learning outcomes. The writing of a bachelor thesis has both a learning functionality, i.e., the writer learns during the writing, and an examination functionality, i.e., the completed thesis represents a measure of the writer's knowledge and skills. An answer to the question certainly depends on the examinee's approach. In the present case, the examinee is considered to be highly ambitious and conscientious. Due to what he perceives as the continuous need for validation, learning functionality is inherent, especially also due to ChatGPT acting as learning partner as described above. The assessment functionality does not seem to be diminished, since due to the accompanying learning processes the bachelor thesis represents a real measure of the knowledge of the examinee. On the other hand, however, less ambitious writers might use ChatGPT for pure structure and text generation, bypassing validation to a large extent, resulting in poor learning outcomes. Then it is up to the conscientiousness of the examiners to what extent the non-validated content of the bachelor thesis is revealed, and the assessment functionality is preserved.

One of the limitations of the study is the small sample size, i.e., only the experiences from writing one bachelor thesis were described. Besides a dependency on the individual commitment of the writer, there might also be a dependency on the topic.

## 6    Conclusions

Digital tools that integrate artificial intelligence are increasingly common in formal education environments. A milestone was the release of the ChatGPT chatbot in November 2022. ChatGPT is based a large language Models (LLM) capable of interpreting and generating text. The free availability of ChatGPT causes especially text heavy teaching formats to be limited in their effectiveness regarding learning and assessment. Accordingly, the question is raised how ChatGPT affects the generation of theses. For clarification, the use of ChatGPT was explicitly allowed for the writing of a bachelor thesis wherever it would be useful. Accompanying this, an autoethnographic log of ChatGPT usage was created. Based on the analysis of this log, we were able to identity various potentials of ChatGPT, such as structuring, brainstorming, and text revision. Among the challenges of its use is the permanent requirement of validation, which at the same time has to be seen as a trigger for learning. Overall, we found that for the case at hand of a research-heavy and highly committed thesis, ChatGPT did not lead to a reduction of thesis-typical learning and assessment functionality. An increase in productivity seems likely. However, due to the expected - and with further versions yet demonstrated - performance improvements of ChatGPT, our findings have limited durability. A re-evaluation should take place at regular intervals to determine the extent to which the assessment functionality is maintained. ChatGPT and, correspondingly, LLMs are to be regarded as valuable digital learning tools, but they require the review and reconceptualization of teaching formats at regular intervals.